\documentclass{cpcauth}
\usepackage{epsfig}
\begin{document}
\begin{frontmatter}
\title{Fluctuations of steps on crystal surfaces}

\author[AC]{W. Selke\corauthref{ws}},
\author[AC,SZ]{F. Szalma}, and
\author[MD]{J. S. Hager}
\address[AC]{Institut f\"ur Theoretische Physik, Technische Hochschule, 
D--52056 Aachen, Germany}
\address[SZ]{Institute for Theoretical Physics, Szeged University, H--6720 Szeged, Hungary}
\address[MD]{Institute of Physical Science and Technology, University of Maryland, College Park, MD, 20742 USA}
\corauth[ws]{e-mail: selke@physik.rwth-aachen.de}
\maketitle
\begin{abstract}

Fluctuations of isolated and pairs of ascending steps of monoatomic height
are studied in the framework of SOS models, using mainly
Monte Carlo techniques. {\it Below} the
roughening transistion of the surface, the profiles of long steps
show the same scaling features for terrace and surface diffusion. For
a pair of short steps, their separation distance
is found to grow as $t^{1/3}$
at late stages. {\it Above} roughening, simulational data on
surface diffusion agree well with the classical continuum theory of Mullins.
\end{abstract}
\begin{keyword}
Monte Carlo simulations \sep SOS model \sep terrace diffusion \sep step
fluctuations \sep surface diffusion
\PACS 05.20.Dd \sep 68.35.Ja \sep 68.35.Rh
\end{keyword}

\end{frontmatter}

\section{Introduction}

In recent years, the dynamics of steps of monoatomic height on crystal
surfaces has attracted much interest, both experimentally and 
theoretically. Experimentally, fluctuations of isolated steps
as well as trains of steps on vicinal surfaces
have been studied thoroughly \cite{jeong}.\\

Theoretically, three distinct mechanisms
driving the step fluctuations have been identified, step
diffusion, evaporation--condensation and terrace
diffusion. Predictions of
Langevin descriptions \cite{bar,bla,kha,pi} have
been checked, confirmed, and extended in Monte Carlo
simulations on discrete SOS
models \cite{bar,bis,szal}. Recently, extensive 
simulations on step diffusion and evaporation--condensation have
been performed, for isolated and pairs of ascending
steps \cite{szal}. Here, results
of a related study on terrace diffusion will be presented.\\

Of course, the roughness of the surface crucially influences
the step dynamics. {\it Below} the roughening transition
temperature, $T < T_R$, the surface is smooth. Terrace diffusion may
now be described in different ways. In an idealized description, the
elementary move consists of the detachment of a step atom, followed
by its random walk on the perfectly flat terrace, and the final
attachment of the atom at the same or neighbouring step. This
description may be simplified further by assuming that the
time of the random walk may be neglected compared to that of
the step processes. More realistically, terrace diffusion
results from jumps of each surface atom to neighbouring sites. This
type of kinetics is usually called surface diffusion, and, in
the following, only the idealized kind of kinetics
will be called terrace diffusion. Obviously, step positions
are uniquely defined for terrace, but not for 
surface diffusion. One of the aims of the present Monte Carlo study
is to compare simulational data, in the framework of SOS models, for
terrace diffusion and surface diffusion.\\ 

{\it Above} the roughening transition, steps are definitely no
longer microscopically well defined; so
the analysis will be restricted to surface diffusion. One
may study the evolution of the profile of initially straight
steps. Indeed, the corresponding equilibration problem, at
$T > T_R$, has been described by Mullins many years
ago \cite{mullins}, and we shall compare our Monte Carlo
findings to that classical theory. Likewise, some of our findings at
$T < T_R$ may be compared not
only to Langevin theories on step fluctuations but also to
theories on the equilibration of surface
profiles \cite{villain,spohn,kandel,bonzel}.\\

The article is organized accordingly, presenting first our
results on isolated and pairs of ascending steps below
roughening, and then those on steps above roughening.

\section{Below roughening}

We simulate square surfaces with isolated
and pairs of steps of monoatomic height. Initially, at time $t$ = 0, the
steps are perfectly straight, and the
bordering terraces are perfectly flat; pairs of steps
are usually separated by at most one lattice spacing. Step
fluctuations result from terrace or surface diffusion.\\

In case of terrace diffusion, the acceptance rates of detaching
and attaching atoms at steps, with a random
walk in between, is assumed to be given by the Boltzmann
factor of the change in the kink
energies as described by the
one--dimensional SOS model. There the kink energy is
proportional to the number of missing bonds to the neighbouring step
sites, i.e. $\epsilon \vert u^s(l) - u^s(l \pm 1) \vert$, $u^s(l)$
being the position of step $s$, ($s$=1, 2), at site $l$. The
time unit, one Monte Carlo step (MCS), is assigned to
$L$ (or 2$L$) attempted elementary moves for isolated (pairs of)
steps of length $L$. In case of pairs of steps no crossing
of steps is allowed.--To speed up simulations, one may replace
the actual random walk by a probability distribution \cite{bla,bis}.\\

In case of surface diffusion, the acceptance rates for jumps of surface
atoms to neighbouring sites will be given by the
Boltzmann factor of the corresponding energy change of the
two--dimensional SOS model, where the local energy is given
by $\epsilon \vert h(i,j) - h(i',j') \vert$, with $(i,j)$ and
$(i',j')$ being neighbouring surface sites. The 
roughening transition is known to occur
at $k_BT_R/\epsilon \approx 1.25$. The time unit, one
MCS, is assigned to $LM$ attempted
jumps, where $L$ is the step length, say, $j= 1,...,L$, and
$M$ refers to the other direction of the surface. To monitor the
step fluctuations, we recorded the step
profile, $z(i,t)$ = $\langle \sum h(i,j) \rangle/L$, summing
over $j$ and averaging over $N$ Monte Carlo realizations with
different random numbers. To 
stabilize the steps, the heights at the boundary lines
parallel to the initial straight steps are kept constant during
the simulation, e.g. for
pairs of steps at $h$=0 and $h$=2.\\      

The step dynamics may be described, both for surface and terrace
diffusion, by the time evolution of the step profile
$z(i,t)$, and, for terrace diffusion, by the average step
positions, $u^s_0(t)$, and the fluctuation function
$w^s(t) = \sqrt{\langle(u^s(l,t) - u^s_0(t)^2) \rangle}$, averaging
over step sites and realizations.\\

\begin{figure}
\centerline{\psfig{figure=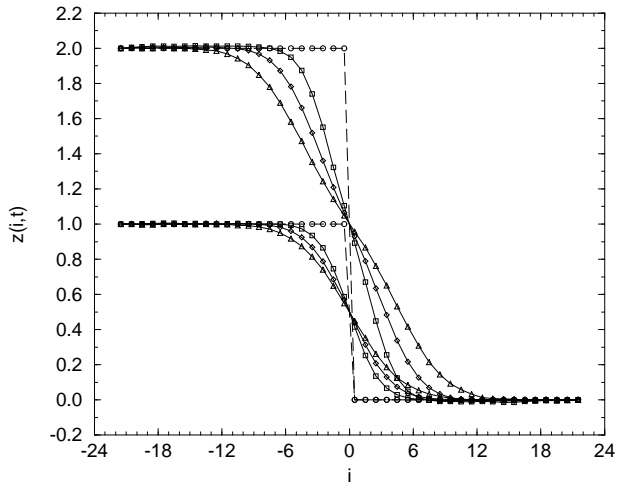,width=9.0cm,angle=0}}
\caption{Simulated step profiles $z(i,t)$ of the two-dimensional SOS model
at $k_BT/\epsilon = 1.0$ with $44 \times 1000$ sites
 at $t$= 0 (circle), 6000 (square), 30000 (diamond), and 120000 (triangle)
MCS, using
surface diffusion. Averages have been taken over 70 realizations.}
\label{fig1}
\end{figure}

Typical step profiles, for isolated and
pairs of steps, are depicted in Fig. 1. To analyse their
scaling behaviour, we use the
ansatz $z(i,t)= z_s(i/t^b)$. At a given height, $z = z_f$, one may yield
at each time, by interpolation, the corresponding
distance $i_f(t)$, and thence, setting
$i_f \propto t^{b_f}$, one may obtain an effective
exponent $b_f(t,z_f)$. Of course, full
scaling holds only when $b_f$ eventually becomes independent of
time $t$ and $z_f$.\\

Indeed, for rather long steps, of at least a few hundred sites, terrace
diffusion and surface diffusion lead to consistent Monte
Carlo results on the effective exponent $b_f$. Most of the simulations
were done at $k_BT/\epsilon$= 1.0 and 0.8 for surface diffusion
and at 1.0 for terrace diffusion.\\

For pairs of steps, in case of
terrace and surface diffusion, the effective exponent $b_f$ is observed
to approach 1/5 at large times of up
to several $10^5$ MCS (well before the step
fluctuations saturate due to the finite step length), when
$z_f$ approaches 0 or 2. For reasons of symmetry, one may
restrict the discussion to $0 < z_f < 1$. At those
times, say, $10^4 < t < 10^6$ (MCS), applying terrace diffusion, the
separation distance between the steps, $d(t) = u_0^1(t) - u_0^2(t)$, and
the fluctuations $w^{1,2}(t)$ follow closely
the power--law $w,d \propto t^{1/5}$. However, in that
time regime, for terrace
and surface diffusion, the effective
exponent $b_f$ changes significantly with $z_f$, increasing with $z_f$ 
from about 1/5 at small heights
rather slowly up to about $0.23$ at $z_f \approx 0.8$, and
then more rapidly to roughly 1/3 as $z_f$ 
approaches 1, depending only weakly on time. It
remains to be seen whether full scaling of
the step profile, with, possibly,  $b= 1/5$, holds in the
limit $L, t \longrightarrow \infty$. Note that the continuum theory
of Spohn on the equilbration of steps due to surface 
diffusion \cite{spohn} does not provide an easy answer to this
question. There the oscillatory character of the profile is
emphasized, which may affect the scaling behaviour. In fact, oscillations
show up, but with very small amplitudes; see Fig. 1 and next
section.-- The value 1/5 for the exponent describing the separation
of the two steps, $d(t)$, has been argued before to follow from the continuum
theory of Rettori and Villain \cite{bar,villain}.\\

Terrace diffusion for pairs of short steps may be described, in the
limit $L \longrightarrow 1$, by two points on a line
emitting particles which execute one--dimensional random walks. When
the emitted particle returns to the emitter, that point will stay
at its original position, while the other point will move by one when the
particle hits that point. The description may also be
applicable to the motion of a pair of kinks along a smooth step in the case of
step--edge diffusion when there are only those two kinks. From our
simulations of such random walks, we infer that
the distance $d$ between the two steps of length $L = 1$ (or the two
kinks) increases as $d \propto t^{1/3}$ at large times.\\

For isolated long steps, the step profiles tend to scale, at sufficiently
large times, with $b \approx 1/6$, for surface diffusion as well as for
terrace diffusion. Indeed, the critical exponent of
the power--law describing the step fluctuations $w(t)$   
at those times is about 1/6 as well, in accordance
with previous simulations for terrace diffusion \cite{bis}. That
result confirms the validity of Langevin descriptions for step
dynamics at late stages \cite{bla,kha}.

\section {Above roughening}

We studied isolated and pairs of ascending steps in the framework of
two--dimensional SOS models at $T > T_R$, or 
one--dimensional SOS models, which are rough at all temperatures
$T > 0$, computing the step profiles $z(i,t)$, applying
surface diffusion. All cases lead to similar results, because above
roughening individual steps are smeared out completely.\\

In particular, the step profiles scale, already at
moderate times of typically a few $10^4$ MCS, with the
critical exponent $b \approx 1/4$, the effective exponent $b_f$
depending only very weakly on $z_f$. The
profiles show oscillations, with the amplitude decreasing rapidly with
distance from the center of the
surface, see Fig. 2. The onset of the oscillations
may be readily understood by calculating the energetics of the
first few excitations, starting from flat terraces and straight steps.
Already the first move leads to an overshooting of the profile at the
next--nearest distance from the center, $i$ =0, which again triggers an
undershooting at further distance, and so on. This effect has been described
before by Mullins in a continuum theory of surface
equilibration above roughening \cite{mullins}. Note that the oscillations
persist to temperatures below roughening, as discussed
above; however, the amplitudes become much smaller, at least at the
times used in our Monte Carlo study, see also Fig. 1.\\

\begin{figure}
\centerline{\psfig{figure=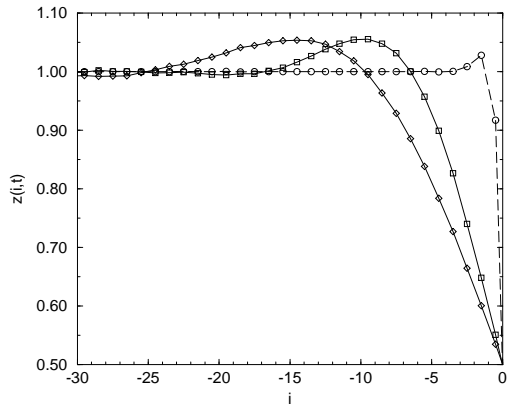,width=7.5cm,angle=0}}
\caption{Part of simulated step profile for isolated steps of 
monoatomic height above roughening. Data for the one--dimensional
SOS model at $k_BT/\epsilon = 1.0$ with 152 sites have been taken
at 1 (circle), 1000 (square) and 5000 (diamond) MCS, averaging
over $10^6$ realizations.}
\label{fig1}
\end{figure}

In Mullins' theory, the basic equation reads    
$dz/dt = -A (d^4 z/d x^4)$, where
$A$ is a temperature dependent coefficient; the continuum
variable $x$ corresponds to $i$ in the discrete description. The equation
may easily be solved by Fourier analysis \cite{mullins}. The resulting
step profiles resemble closely those found in the
simulations; differences are expected to show up only
at early stages
of equilibration, as observed before for other surface
defects such as periodic grooves \cite{sel2}. From
the basic equation, it follows that the amplitudes of the oscillations
settle at fixed values, independent of 
temperature. We confirmed these features by simulating steps
at various temperatures. Actually, the continuum description
of Mullins leads to a perfect scaling of the step profiles with
$b = 1/4$.\\
  
F. Sz. thanks the
Hungarian National Research Fund, under grant number OTKA D32835,
and the Deutsche Forschungsgemeinschaft
for financial support.

\end{document}